\documentclass[twocolumn,final,aps,prl,showpacs,superscriptaddress,amsmath,amssymb,amsfonts,floatfix,showkeys]{revtex4-1}

\usepackage{graphicx}
\usepackage{booktabs}
\usepackage{multirow}
\usepackage{makecell} 
\usepackage{array}
\usepackage{url}
\usepackage{amsmath}   
\usepackage{siunitx}  
\usepackage{float}
\usepackage[normalem]{ulem}

\usepackage[colorlinks,breaklinks,bookmarks=true,citecolor=blue,linkcolor=blue,urlcolor=blue]{hyperref}
\usepackage{color}
\usepackage{tabularx}
\usepackage{sidecap}
\usepackage{orcidlink}
\usepackage{soul}

\newcolumntype{Y}{>{\centering\arraybackslash}X}
\begin{document}

\author{Ao Zhang\orcidlink{0000-0001-9786-3942}}
\affiliation{School of Physics, Northwest University, Xi'an 710127, China}
\affiliation{Department of Physics, Southeast University, Nanjing 211189, China}

\author{Chao Deng\orcidlink{0009-0003-8433-2909}}
\affiliation{School of Physics, Northwest University, Xi'an 710127, China}

\author{Leonard M. Verhoff\,\orcidlink{0009-0004-3358-1312}}
\affiliation{Institute of Solid State Physics, TU Wien, 1040 Vienna, Austria}

\author{Mi Jiang\orcidlink{0000-0002-9500-202X}}
\affiliation{School of Physical Science and Technology, Soochow University, Suzhou 215006, China}
\affiliation{Jiangsu Key Laboratory of Frontier Material Physics and Devices, Soochow University, Suzhou 215006, China}

\author{Run Yang\orcidlink{0000-0002-9547-756X}}
\affiliation{Department of Physics, Southeast University, Nanjing 211189, China}
\affiliation{Key Laboratory of Quantum Materials and Devices of Ministry of Education, Department of Physics, Southeast University, Nanjing 211189, China}

\author{Zhixiang Shi\orcidlink{0000-0003-3881-5152}}
\email{zxshi@seu.edu.cn}
\affiliation{Department of Physics, Southeast University, Nanjing 211189, China}
\affiliation{Key Laboratory of Quantum Materials and Devices of Ministry of Education, Department of Physics, Southeast University, Nanjing 211189, China}

\author{Karsten Held\,\orcidlink{0000-0001-5984-8549}}
\email{held@ifp.tuwien.ac.at}
\affiliation{Institute of Solid State Physics, TU Wien, 1040 Vienna, Austria}

\author{Liang Si\orcidlink{0000-0003-4709-6882}}
\email{siliang@nwu.edu.cn}
\affiliation{School of Physics, Northwest University, Xi'an 710127, China}
\affiliation{Shaanxi Key Laboratory for Theoretical Physics Frontiers, Xi'an 710127, China}
\affiliation{Fundamental Discipline Research Center for Quantum Science and technology of Shaanxi Province, Xi'an 710127, China}
\affiliation{Institute of Solid State Physics, TU Wien, 1040 Vienna, Austria}

\title{Reconciling Models for Infinite-Layer Nickelate Superconductors: Role of Interstitial-$s$ Band or $A$-Pocket}

\begin{abstract}
Understanding the minimal model that captures the essential physics of infinite-layer nickelate superconductors remains a central challenge. 
Using first-principles density-functional theory combined with dynamical mean-field theory (DFT+DMFT), we compute and compare the strongly correlated electronic structures and optical spectra of various effective models for the infinite-layer nickelate LaNiO$_2$, ranging from a $d_{x^2-y^2}$ 1-band model to a full 6-band model with all Ni-$3d$ plus an interstitial-$s$ orbital. The results demonstrate that Ni-$3d$ 5-band models overestimate the Hundness picture, whereas 6-band models, which explicitly include the interstitial-$s$ band and have an $A$-pocket, recover the characteristic Mottness physics. Importantly, the influence of the interstitial-$s$ band on the correlations of the $d_{x^2-y^2}$ band is negligible. 
Thus, superconductivity in infinite-layer nickelates can be reliably described within a single-band Hubbard framework, while the accurate description of inter-orbital properties such as (optical) conductivities  requires the inclusion of at least the interstitial-$s$ band.
\end{abstract}


\maketitle

\emph{Introduction---}The discovery of superconductivity in infinite-layer nickelates \cite{li2019superconductivity,osada2021nickelate,hepting2020electronic,gu2020single,zeng2022superconductivity,lee2023linear,PhysRevX.15.021048,chow2025bulk} has sparked the study of unconventional superconductors and stimulated efforts to understand the minimal effective model of this new family of superconductors \cite{PhysRevLett.125.077003,kitatani2020nickelate,PhysRevResearch.2.013214,PhysRevB.101.075107,PhysRevLett.125.077003,PhysRevB.101.041104,PhysRevResearch.1.032046,PhysRevB.102.100501,PhysRevB.108.245115,PhysRevB.105.054516,PhysRevB.84.165119,PhysRevX.10.011024,PhysRevB.101.081110,PhysRevX.10.041002}. 
In cuprates, the nominal $\mathrm{Cu}^{2+}\,(3d^{9})$ configuration, mediated by strong hybridization between Cu-3$d$ and O-2$p$ orbitals, gives rise to a single half-filled Zhang-Rice singlet \cite{PhysRevB.37.3759} and the effective two-dimensional Emery model \cite{PhysRevLett.58.2794} is the minimal model. Upon hole or electron doping, this correlated Mott insulator evolves into a high-$T_C$ superconducting phase, a behavior captured qualitatively by minimal single-band models despite the complexity of their multi-orbital origins.
Like cuprate superconductors \cite{bednorz1986possible,anderson1997theory,scalapino1995case,RevModPhys.72.969}, the infinite-layer nickelates possess a layered NiO$_2$ square lattice of Ni ions embedded in O ligands, suggesting that strong correlation physics from the partially filled Ni-3$d$ underlies their normal and superconducting states under hole doping induced by Sr or Ca substitution. 

By analogy, early theoretical \cite{kitatani2020nickelate} and experimental \cite{cheng2024evidence} studies of infinite-layer nickelates posited that a single $d_{x^2-y^2}$ band dominating the Fermi surface could similarly host strong correlations and cuprate-like unconventional superconductivity. Experimental evidence from transport \cite{ding2024cuprate} and spectroscopic measurements \cite{cheng2024evidence,PhysRevB.111.014519} supports key aspects of this picture, including signatures of strong electron correlations and a cuprate-like dome-shaped superconducting region \cite{PhysRevLett.125.147003,PhysRevLett.125.027001}. However, recent high-resolution angle-resolved photoemission spectroscopy (ARPES) experiments \cite{tptb-8hb4} have also
revealed a pocket near the $A$-momentum, besides the 
Ni $d_{x^2-y^2}$  Fermi surface.
At the $A$-point itself, this band consists of a hybridization between Nd‑5$d$ and Ni‑3$d$ orbitals.
At higher energy also
a spatially delocalized interstitial‑$s$ band 
contributes substantially. Altogether we have 
five Ni-3$d$, five Nd-5$d$ and one 
interstitial‑$s$ band around the Fermi energy, with the Ni-$d_{x^2-y^2}$ and $A$-pocket
crossing the Fermi energy in ARPES \cite{tptb-8hb4}.
The simplest Hamiltonian modelling this consists of one or more Ni-3$d$ orbitals plus a single extra orbital for the $A$-pocket.
Despite its mixed nature, this single extra orbital is then typically referred to
as interstitial $s$-orbital  in the literature \cite{PhysRevB.100.205138,gu2020substantial,PhysRevB.102.100501} and 
also henceforth.
The presence of this interstitial-$s$ orbital complicates the identification of the truly (low-energy) minimal effective model and raises questions about the relative roles of Mottness 
\cite{PhysRevB.101.020501,PhysRevLett.109.156402,goodge2021doping,kitatani2020nickelate,PhysRevResearch.4.043036} 
vs.~Hundness \cite{PhysRevB.101.041104,kang2023infinite,PhysRevLett.126.127401} physics in nickelates.

In this Letter, we employ first-principles density-functional theory \cite{PhysRev.136.B864,PhysRev.140.A1133} combined with dynamical mean-field theory (DFT+DMFT) 
\cite{PhysRevLett.62.324,RevModPhys.68.13,kotliar2004strongly,held2007electronic} to systematically compare the strongly correlated electronic structures and optical responses of multiple effective models for infinite-layer nickelate LaNiO$_2$. 
Specifically, we construct 1-band ($d_{x^2-y^2}$), 2-band ($d_{x^2-y^2}$+interstitial-$s$ band), 5-band (Ni-3$d$), and 6-band models (Ni-3$d$+interstitial-$s$) derived from maximally localized Wannier functions \cite{RevModPhys.84.1419}, and calculate their frequency- and momentum-dependent spectral functions, correlations, and optical conductivities.

Our multiple model comparison reveals that 5-band models without the interstitial-$s$ orbital tend to overestimate Hundness character, leading to reduced quasi-particle coherence, multi-orbital behavior and suppressed spin correlations. In contrast, 6-band models that explicitly include the interstitial-$s$ band restore Mott-characteristic physics, evidenced by the enhanced quasi-particle coherence, orbital occupation and strong mass renormalization of the Ni-3$d_{x^2-y^2}$ orbital.

However, if the doping of the  Ni-3$d_{x^2-y^2}$ orbital is adjusted
to account for the self doping 
induced by the interstitial-$s$ orbital,
the 1-band Hubbard and 5-band Ni-3$d$ models show very similar correlations of the $d_{x^2-y^2}$   orbital as the 6-band model which explicitly includes
the interstitial-$s$ orbital.
This reconciles the experimental observation of an interstitial-$s$  orbital/$A$-pocket at the Fermi energy (E$_f$) with the success of the minimal $d_{x^2-y^2}$ description in prior theoretical calculations for superconductivity. For inter-orbital transitions and optical responses, on the other hand, the inclusion of the interstitial-$s$ band is absolutely necessary. Our results thus provide a unified perspective on the appropriate minimal model for infinite-layer nickelates and establish a roadmap for explaining future theoretical and experimental studies.

\begin{figure*}[!htbp]
\centering
\includegraphics[width=1 \textwidth]{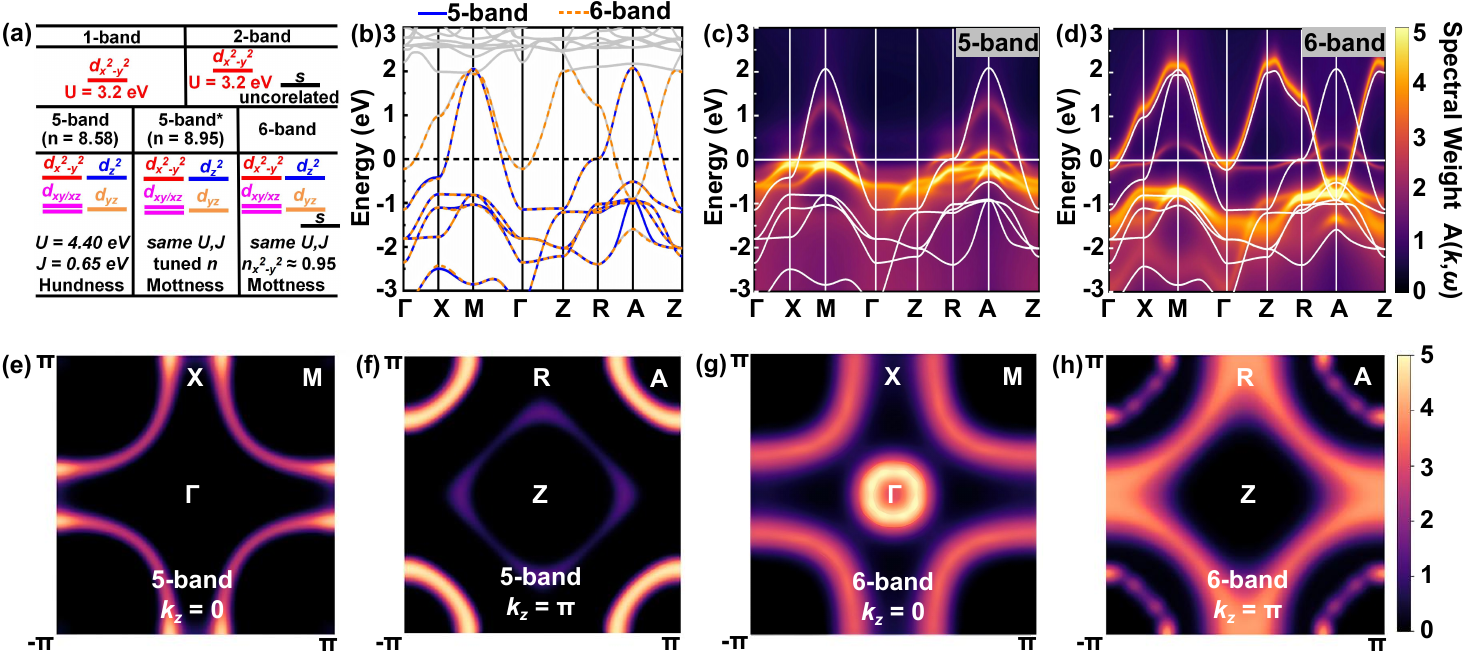}
\caption{(a) Schematic illustration of the different models, where the interstitial-$s$ band is approximately treated as uncorrelated, and only for 5- and 6-band models a multi-band setup of interaction parameters with $U$, $J$, and $U'$ is employed. (b) DFT bands vs.~Wannier projections of 5- and 6-band models. (c,d) DFT (Wannier) bands vs.~DMFT spectral functions $A$($\omega$,$k$) for 5- and 6-band models. (e-h) DMFT-computed Fermi surfaces for (e,f) 5- and (g,h) 6-band models at (e,g) $k_z$=0 and (f,h) $k_z$=$\pi$, respectively.}

\label{fig-electronic}
\end{figure*}

\emph{Method---}DFT-level first-principles electronic structures were computed using the \verb|Vasp| \cite{Kresse1996PRB,Kresse1996CMS} and \verb|Wien2K| \cite{Blaha2001WIEN2k,schwarz2002electronic} codes. The exchange-correlation functional was treated with the Perdew-Burke-Ernzerhof version of the generalized gradient approximation (GGA-PBE) \cite{Perdew1996PRL}. The lattice parameters of $a$=$b$=3.89\,\AA ~and $c$=3.34\,\AA~are from Ref.~\cite{si2024closing}. Brillouin-zone integrations were carried out on a 13$\times$13$\times$15 $k$-mesh. 
The corresponding DFT bands were projected onto maximally localized Wannier functions \cite{RevModPhys.84.1419} to construct the low-energy effective Hamiltonians using the \verb|Wannier90| code \cite{mostofi2008wannier90} and \verb|Wien2Wannier| interface \cite{kunevs2010wien2wannier}.
As input for the DMFT calculations, five different (low-energy) models were constructed in the vicinity of the E$_f$, see Fig.~\ref{fig-electronic}~(a): (1) 1-band model containing the $d_{x^2-y^2}$ band; (2) 2-band model including the $d_{x^2-y^2}$ and the interstitial-$s$ bands; (3) 5-band model including the Ni-$3d$ bands; (4) 5-band$^*$ model which is the same as (3) except for an adjusted doping
to account for the number of electrons in the (otherwise not further included) interstitial-$s$ band, and (5) 6-band model including the Ni-$3d$ plus the interstitial-$s$ band.

The resulting tight-binding Hamiltonians were supplemented with the local Kanamori interactions, and the fully localized limit was employed for the double-counting correction \cite{liechtenstein1995density,anisimov1993density}. The interaction parameters were chosen following commonly used values in the literature Refs.~\cite{PhysRevLett.130.166002,PhysRevLett.124.166402}: $U(d_{x^2-y^2})$=3.2\,eV for the 1-band model; the interstitial-$s$ band was treated as uncorrelated; finally, when including all five Ni-$3d$ orbitals, the interaction parameters were set to intra-orbital $U$=4.4\,eV, Hund’s exchange $J$=0.65\,eV, and inter-orbital interaction $U'$=$U$-2$J$=3.1\,eV. The resulting many-body Hamiltonians were solved at room temperature (300\,K) within the DMFT framework using the continuous-time quantum Monte Carlo (\verb|CT-QMC|) impurity solver in the hybridization expansion \cite{gull2011continuous} as implemented in the \verb|W2dynamics| code \cite{parragh2012conserved,wallerberger2019w2dynamics}. Real-frequency spectral functions were obtained via analytic continuation using the maximum entropy method (\verb|MaxEnt|) \cite{sandvik1998stochastic, gubernatis1991quantum}, as implemented in the \verb|ana_cont| program~\cite{kaufmann2023ana_cont}.
The (uncorrelated) DFT-level and the (correlated) DMFT-level optical conductivities of LaNiO$_2$ were computed using \verb|Wien2K| \cite{Blaha2001WIEN2k, ambrosch2006linear} and the \verb|WOPTIC| code \cite{assmann2016woptic}, respectively.

\emph{Model dependent electronic structure---}The differences between different models in DMFT calculations can be traced back to the numbers of correlated and uncorrelated bands.
Figure~\ref{fig-electronic}(a) provides a schematic overview of several low-energy models, including the Ni-3$d_{x^2-y^2}$ 1-band model \cite{kitatani2020nickelate,PhysRevX.10.021061,klett2022magnetic}, 2-band model ($d_{x^2-y^2}$+interstitial-$s$) \cite{PhysRevB.108.245115,PhysRevResearch.1.032046,PhysRevB.106.134504,PhysRevB.102.100501,plienbumrung2023single,nhg9-l5l1,PhysRevB.105.054516}, 5-band model (Ni-3$d$) \cite{PhysRevLett.126.127401}, and 6-band model (Ni-3$d$+interstitial-$s$) \cite{PhysRevLett.125.077003}.
We also consider a 5-band$^*$ model (see Table~\ref{tab:hopping}) with an adjusted filling of $n$=8.95, in which the chemical potential is tuned so that the $d_{x^2-y^2}$ orbital occupation satisfies $n_{x^{2}-y^{2}}$$\approx$0.95 as computed in Ref.~\cite{hausoel2025superconducting}.
As discussed above this way the self-doping due to the interstitial-$s$ orbital is accounted for. And, as discussed below, this adjustment restores the Mottness character in the Ni-3$d$ 5-band model.

Figure~\ref{fig-electronic}(b) compares the Wannier-derived tight-binding bands of the 5- and 6-band models with the DFT band structure of LaNiO$_2$; for the Wannier projections of the other models see Supplementary Materials (SM) Section I \cite{SM}. Both models accurately reproduce the DFT bands along the high-symmetry 
$k$-paths, demonstrating that the constructed tight-binding Hamiltonians provide an effective low-energy description. The only notable discrepancy appears in the 5-band model: at the $A$-point, between -2\,eV and -1\,eV, the Wannier bands deviate from the DFT ones. This mismatch arises because the band at the $A$-point around -1.8\,eV lacks Ni-3$d$ character and is therefore not captured by the 5-band (Wannier) projection (for the LaNiO$_2$ DFT band characters, see SM Section II \cite{SM}). This mismatch is resolved in the 6-band projection, which includes the additional interstitial-$s$ orbital (i.e., the pocket band at the $A$- and $\Gamma-$point), highlighting the necessity of including the interstitial-$s$ (or pocket) band when addressing multi-band effects in nickelates. It should be noted that the contribution of the interstitial-$s$ orbital to the low-energy electronic structure exhibits a pronounced momentum dependence. 
To study the origin and character of the pocket band, we plot its real-space, momentum-resolved charge distribution in SM Section III \cite{SM}. At the $\Gamma$-point, the interstitial-$s$ band is dominated by hybridized La-5$d_{z^2}$ and Ni-3$d_{z^2}$ states, whereas near the $A$-point it includes a sizable contribution from  an $s$-like orbital localized above the Ni sites.

\begin{table*}[t]
\caption{Model definitions, Wannier hopping parameters [along (000: on-site hopping), (100: first-nearest), (110: 2nd-nearest), (200: 3rd-nearest), and (001: out-of-plane) directions] of Ni-3$d_{x^2-y^2}$ (in units of meV), interaction parameters, total and $d_{x^2-y^2}$ orbital occupancies in Wannier projection and DFT+DMFT calculations. The 5-band$^*$ model employs the same Hamiltonian and interaction parameters as the 5-band model, with only the electron filling adjusted (for details see SM Section IV \cite{SM}).} 
\label{tab:hopping}
\centering

\begin{tabular}{c|c|c|c|c|c|c|c|c|c|c|c|c}
\hline
\hline
Model
& {Orbital(s)}
& {$t_{000}$}
& {$t_{100}$}
& {$t_{001}$}
& {$t_{110}$}
& {$t_{200}$}
& {$t_{110}/t_{100}$}
& {$t_{200}/t_{100}$}
& {$U$ (eV)}
& {$J$ (eV)}
& {$n_{total}$}
& {$n_{d_{x^2-y^2}}$} \\
\hline
1-band & $d_{x^2-y^2}$ & 262.5 & -385.5 & -35.1 & 96.1 & -45.0 & -0.249 & 0.117 & 3.20 & 0.00 & 0.95 & 0.95 \\
2-band & $d_{x^2-y^2}$+interstitial-$s$ & 289.5 & -346.0 & -31.0 & 72.9 & -47.5 & -0.211 & 0.137 & 3.20 & 0.00 & 1.00 & 0.92 \\
5-band & Ni-$3d$ & 257.0 & -385.6 & -37.3 & 97.7 & -45.9 & -0.253 & 0.119 & 4.40 & 0.65 & 8.58 & 0.75 \\
~5-band* & Ni-$3d$ & 257.0 & -385.6 & -37.3 & 97.7 & -45.9 & -0.253 & 0.119 & 4.40 & 0.65 & 8.95 & 0.95 \\
6-band & Ni-$3d$ + interstitial-$s$ & 262.5 & -383.6 & -37.8 & 95.6 & -44.9 & -0.249 & 0.117 & 4.40 & 0.65 & 9.00 & 0.95 \\
10-band & La-$5d$ + Ni-$3d$ & 281.4 & -391.5 & -52.2 & 97.4 & -48.9 & -0.249 & 0.125 & 4.40 & 0.65 & 9.00 & 0.97 \\

11-band & La-$5d$ + Ni-$3d$ + interstitial-$s$ & 389.0 & -390.2 & -1.2 & 45.1 & -22.2 & -0.116 & 0.057 & 4.40 & 0.65 & 9.00 & 0.93 \\

\hline
\hline
\end{tabular}
\end{table*}

To further examine the electronic structure, particularly the low-energy behavior near the E$_f$, we present in Figs.~\ref{fig-electronic}(c,d) the momentum-resolved spectral functions $A$($\omega$,$k$) of the 5- and 6-band models. In both cases, the $d_{x^2-y^2}$-derived states at $k_z$=0 are well captured [Figs.~\ref{fig-electronic}(e,g)]. However, important differences emerge depending on whether the 
interstitial-$s$ band  is included or not.

First, at $k_z$=0, the pocket at $\Gamma$-point [Fig.~\ref{fig-electronic}(d,g)], which is contributed by the interstitial-$s$ orbital, cannot be reproduced within the 5-band model [Fig.~\ref{fig-electronic}(c,e)], demonstrating that the $\Gamma$-centered pocket has a
clear 3D character rather than the quasi-2D extension of $d_{x^2-y^2}$ orbital. 
For the DMFT spectral functions at 20\% hole doping and with an additional interaction $U$ for the interstitial-$s$ orbital, see SM Section V \cite{SM}.
Our preliminary work \cite{PhysRevLett.124.166402,held2022phase,kitatani2020nickelate} demonstrated that including correlations on the La-5$d$ pocket bands ($U$=2.5\,eV and $J$=0.25\,eV) shifts the $\Gamma$-pocket band upward in energy, such that it may no longer cross the E$_f$. This treatment should be distinguished from the interaction $U_s$ applied to the effective interstitial-$s$ orbital in our 6-band model. For the parameters examined here, the $\Gamma$-pocket persists at $n$=9.0 for both values of $U_s$, whereas it is absent at $n$=8.8 in both cases, for details see SM Section~V \cite{SM}. The hole-doped results ($n$=8.8 for both with and without $U_s$) are consistent with the ARPES observations of Sun \textit{et al.}~\cite{sun2025electronic} for the hole-doped compound.
Nevertheless, the interstitial-$s$ band remains essential for capturing the $A$-pocket and key three-dimensional features of the electronic structure and the relevant inter-band transitions. Therefore, it cannot be neglected in discussions of the \(k_z\)-dependent Fermi surface, as well as transport and optical responses.

Second, the 5-band model exhibits a weaker quasi-particle renormalization and sheet-like Fermi surface of the $d_{x^2-y^2}$ band, despite identical interaction parameters in both models. This is accompanied by a smaller $d_{x^2-y^2}$ orbital occupancy (see Table~\ref{tab:hopping} 5-band: $n_{d_{x^2-y^2}}$=0.75). 

Third, although both models capture the $A$-pocket [see Fig.~\ref{fig-electronic}(f,h)], its physical origin differs: in the 5-band model it is a hole-like pocket, arising from the partially filled Ni-3$d_{xy}$ orbital [Fig.~\ref{fig-electronic}(f)], whereas in the 6-band model it is an electron-like pocket, formed by the (uncorrelated) interstitial-$s$ band.
Specifically, when the Ni-3$d$ orbitals have the same total filling in 5- and 6-band models, i.e., $n$=8.58 for 5-band and $n$=9.0 for 6-band, the noninteracting (i.e., DFT) calculations for 5-band model yield a nearly fully occupied Ni-3$d_{xy}$ orbital ($n$$\sim$2.0), with its band maximum near $A$-point lying below the E$_f$ [see the white lines in Fig.~\ref{fig-electronic}(c)]. Orbital-dependent self-energy corrections, together with the self-consistent adjustment of the chemical potential in DMFT, shift the bands across the E$_f$ near $A$, producing a hole-like pocket [see Fig.~\ref{fig-electronic}(c,f)]. This change is accompanied by a reduction of the $d_{xy}$ occupation from $\sim$2.00 to 1.87 and an increase of the $d_{x^2-y^2}$ occupation from approximately 0.59 to 0.75 (the total filling remains fixed at $n$=8.58).
In contrast, the electron-like $A$-pocket associated with the interstitial-$s$ band is already present in the noninteracting (DFT) 6-band calculation and is retained after including local correlations in DMFT. DMFT therefore describes the correlation-induced renormalization and self-consistent charge redistribution in 6-band model, rather than the emergence of this pocket. In the 6-band model, the interstitial-$s$ occupation decreases from 0.42 (DFT) to 0.35 (DMFT), while the Ni-3$d_{xy}$ occupation remains $\sim$2.0. The $d_{xy}$ branch near $A$ remains below the E$_f$, and hence the corresponding hole-like $d_{xy}$ pocket found in 5-band model is absent.
Finally, the 6-band model reproduces the main features observed in ARPES measurements \cite{tptb-8hb4}, especially at $k_z$=$\pi$ [Fig.~\ref{fig-electronic}(h)].

Table~\ref{tab:hopping} (and SM Section VI \cite{SM}) summarizes the Wannier-projected hopping parameters and electron fillings for the different models considered in Fig.~\ref{fig-electronic}(a). Notably, the single-band projection quantitatively reproduces those obtained from the 5- and 6-band projection. This is a first indication that, once the electron filling is properly adjusted, the superconducting properties of the system can be reliably captured within an effective single-band description, consistent with Refs.~\cite{kitatani2020nickelate,PhysRevLett.130.166002,PhysRevMaterials.6.L091801,PhysRevB.109.235126}. 

\begin{figure}[t]
\centering
\includegraphics[width=\columnwidth,keepaspectratio]{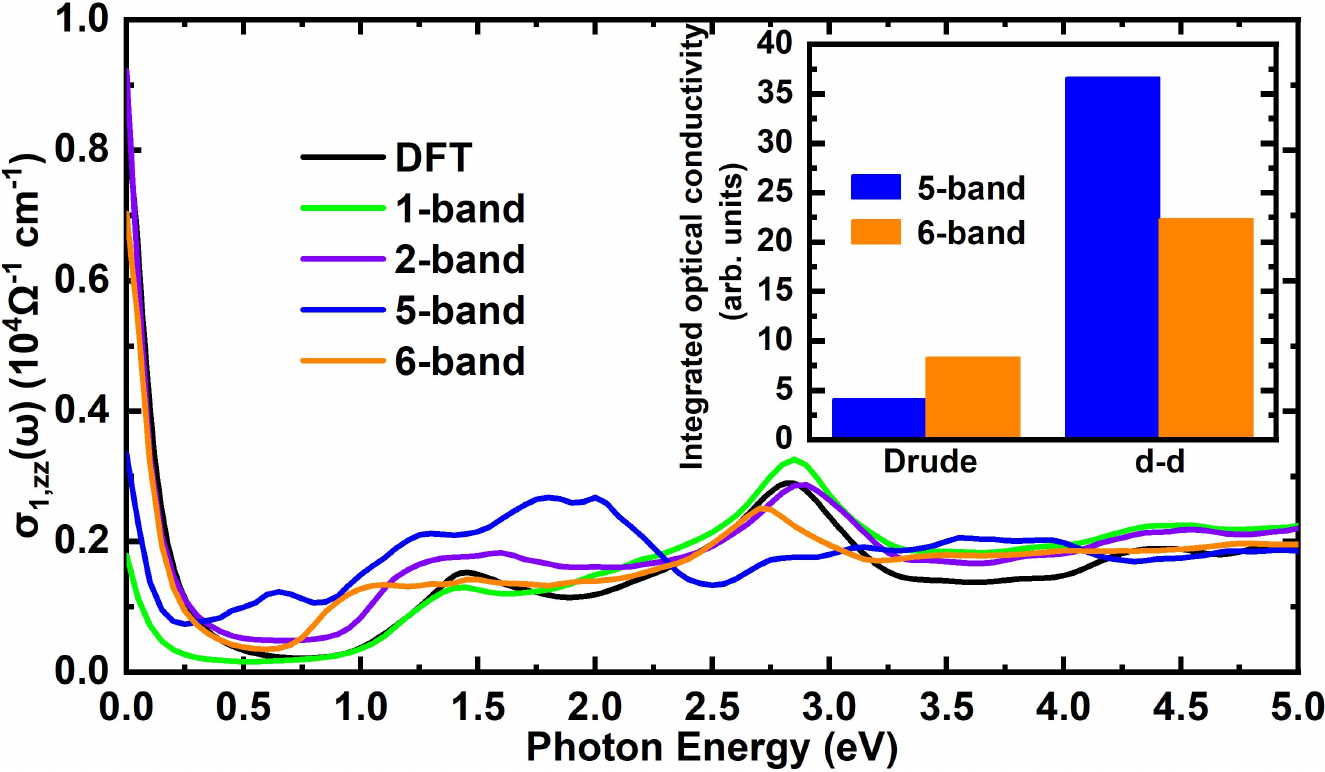}
\caption{DMFT-computed optical conductivity $\sigma$($\omega$) of LaNiO$_2$ for the 1-, 2-, 5-, and 6-band models, as summarized in Fig.~\ref{fig-electronic} and Table~\ref{tab:hopping}, the DFT result is included for reference. The inset displays the integrated optical conductivity over selected frequency windows, where the low-frequency Drude weight is defined as $W_{\mathrm{Drude}}$=$\int_{0}^{\omega_c}\sigma_1(\omega)\,d\omega$ with $\omega_c$=0.296\,eV chosen to exclude inter-band contributions, and the mid-infrared spectral weight is defined as $W_{d\text{-}d}$=$\int_{\omega_1}^{\omega_2}\sigma_1(\omega)\,d\omega$, which characterizes the strength of transitions between multiple Ni-3$d$ orbitals.}
  \label{fig:figure2}
\end{figure}

\emph{Optical conductivity---}Figure~\ref{fig:figure2} presents the out-of-plane optical conductivity 
$\sigma$($\omega$) of LaNiO$_2$ calculated within DMFT \cite{assmann2016woptic} using different models, for the 5-band$^\ast$ model see SM \cite{SM} Fig.~S.4. For comparison, the (non-interacting)
DFT $\sigma$($\omega$) is also shown. In the low-frequency limit ($\omega$$\approx$0\,eV), pronounced differences emerge in the Drude response among the models. The 1- and 5-band models yield a 
relatively weak Drude weight, whereas the 2- and 6-band models exhibit a substantially enhanced Drude response. This highlights the crucial role of the 
interstitial-$s$ band: it adds the $A$-pocket to the Fermi surface which provides an {\it additional} contribution to the Drude weight.
In contrast, the similarity of the Drude peaks between DFT, 2- and 6-band results reflects the weakly correlated or nearly uncorrelated nature of the $A$-pocket contribution of the interstitial-$s$ band.

With increasing frequency $\omega$, the optical responses of the different models deviate in the mid-infrared regime of approximately 0.5\,eV to 1.6\,eV. In particular, the 5-band model shows a pronounced absorption peak near 2\,eV, which cannot be understood in terms of a single-particle picture of inter-band transitions among Ni-$3d$ states [see Fig.~\ref{fig-electronic}(c)]. This feature is absent in the 6-band model. Moreover, all other models, including the DFT, display similar behavior in the higher-energy range from $\sim$1.5\,eV to 5\,eV, indicating that the 5-band model overestimates correlation-induced optical excitations within the multi-orbital Ni-$3d$ manifold near 2\,eV and underscoring the weakly correlated character of the Ni-3$d$ $t_{2g}$, Ni-3$d_{z^2}$ 

and interstitial-$s$ bands. These observations support a picture in which infinite-layer nickelates consist of a strongly correlated single $d_{x^2-y^2}$ band coexisting with weakly correlated (or uncorrelated) auxiliary bands. In both the 2- and 6-band models, the 3D (itinerant) pocket band with dominant interstitial-$s$ orbital character serves as an additional channel for the Drude response. Additionally, it is worth noting that the 6-band model exhibits a pronounced peak $\sim$1\,eV, reminiscent of the correlation-induced feature observed in SrVO$_3$ at $\omega$$\sim$4\,eV \cite{assmann2016woptic}. Based on the above analysis and the DMFT $A$($\omega$,$k$) [Fig.~\ref{fig-electronic}(d)], this peak can be attributed to inter-band transitions between the interstitial-$s$ and Ni-3$d$ orbitals. This explains why the corresponding feature is absent in models that do not explicitly include the interstitial-$s$ band.

To quantify the redistribution of optical spectral weight, the inset of Fig.~\ref{fig:figure2} shows the 
$\sigma$ integrated over selected frequency windows. Compared with the 5-band model, the 6-band model exhibits a significantly enhanced low-frequency spectral weight, accompanied by a corresponding reduction in the intermediate-energy range. Notably, the 2-band model displays a similar enhancement (for computational details and additional results see SM Section VII \cite{SM}) of low-frequency spectral weight, reinforcing the conclusion that the inclusion of an interstitial-$s$ band substantially increases both the intra- and inter-band contribution.

\begin{figure}[t]
\centering
\includegraphics[width=\columnwidth]{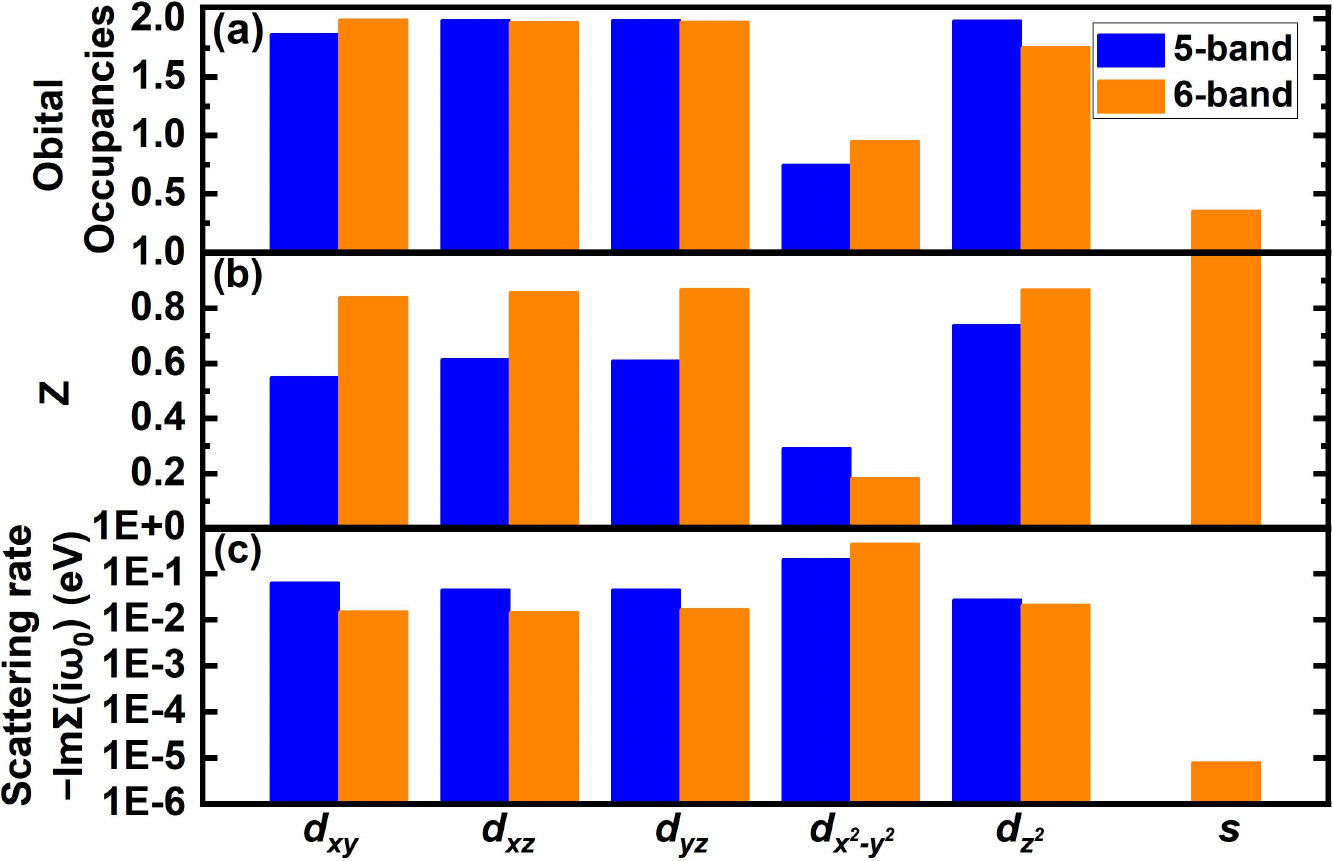}
\caption{DMFT-computed (a) orbital occupation, (b) quasi-particle renormalization factor $Z$, and (c) scattering rate [-$\Im \Sigma$($i\omega$)] for 5- and 6-band models.}  \label{fig:figure3}
\end{figure}

\emph{Hund versus orbital-selective correlations}---Figure~\ref{fig:figure3}(a) shows the DMFT-calculated orbital occupations, comparing the 5- and 6-band models; for the 5-band$^\ast$ model see SM \cite{SM} Fig.~S.5. A key difference is obviously  the interstitial-$s$ band occupation: $\sim$0.35, implying a total Ni-3$d$ occupation of 8.65 electrons in the 6-band model, close to the value in Ref.~\cite{PhysRevLett.126.127401}: 8.58. As a consequence, the occupation of $d_{x^2-y^2}$ orbital differs substantially between the two models: 0.75 in 5-band model and 0.95 in 6-band model, corresponding to a 0.2 electron difference. Such a discrepancy is significant and cannot be neglected in calculations of superconducting properties. In addition, the 5-band model overestimates the occupation of the $d_{z^2}$ orbital due to hybridization 
at the $\Gamma$-point between Ni-3$d_{z^2}$ and La-5$d_{z^2}$ orbitals \cite{hausoel2025superconducting} (this $\Gamma$-pocket is included in the 6-band model through the interstitial-$s$ band).  These results highlight the essential role of the interstitial-$s$ band in correctly redistributing electron occupation.

Figure~\ref{fig:figure3}(b) compares the quasi-particle renormalization factor $Z$ of each orbital in 5- and 6-band models, where $Z^{-1}$=1-$\left.\partial_{\omega}\operatorname{Re}\Sigma(\omega)\right|_{\omega=0}\simeq m^{*}/m_e$, with $m^{*}$ and $m_e$ denoting the DMFT and DFT effective masses, respectively. In the 5-band model, the $Z$ factor for Ni-3$d_{t_{2g}}$, $d_{z^2}$ and $d_{x^2-y^2}$ is roughly 0.59, 0.74, and 0.29, respectively, indicating substantial multi-orbital correlations and explaining the contribution of $t_{2g}$ states to the Fermi surface [Fig.~\ref{fig-electronic}(f) around the $A$-point]. By contrast, the 6-band model shows a much stronger orbital dependence: $t_{2g}$, $d_{z^2}$ and $d_{x^2-y^2}$ orbitals have $Z$ of 0.86, 0.87, and 0.18, respectively. By construction, the uncorrelated $s$-band has $Z$$\sim$1. That is, only the $d_{x^2-y^2}$ orbital is strongly correlated. These results are fully consistent with the optical conductivity analysis in Fig.~\ref{fig:figure2}. Differences in electron filling must be taken into account when comparing with ARPES measurements. Sun \textit{et al.}~reported a low-energy band renormalization up to $Z^{-1}$$\sim$3 in hole-doped La$_{0.8}$Sr$_{0.2}$NiO$_2$ \cite{sun2025electronic}, while their accompanying DMFT calculation yielded a mass enhancement of $\sim$3.0. Using the same real-frequency self-energy derivative method, we find that $Z^{-1}$ for Ni-3$d_{x^2-y^2}$ orbital in the 6-band model decreases from approximately 5.44 at $n$=9.0 to 4.05 at $n$=8.8. Applying the low-frequency Matsubara self-energy fitting procedure gives corresponding values of $\sim$5.39 and 3.30, respectively; corresponding details in SM \cite{SM} indicate that hole doping reduces quasiparticle renormalization and brings the calculated results closer to experiment (see SM Sec.~VIII).

Further insight is provided by the orbital-resolved scattering rates of -$\Im \Sigma$($i\omega_n$) as shown in Fig.~\ref{fig:figure3}(c). In the 5-band model, all Ni-3$d$ orbitals exhibit relatively large and similar scattering rates, indicating short carrier lifetimes. In contrast, the 6-band model shows a pronounced enhancement of the scattering rate only for the $d_{x^2-y^2}$ orbital, while the remaining $d$-bands and the $s$-band exhibit much smaller scattering rates. Taken together, the factor $Z$ and scattering rates reveal fundamentally different correlation regimes in the two models. The 5-band model exhibits collective multi-orbital correlations dominated by Hund’s coupling, with the several $d$-orbitals strongly renormalized. In contrast, the 6-band model has strong correlations primarily in the $d_{x^2-y^2}$ orbital, while the other orbitals remain weakly correlated, indicating  an orbital-selective Mottness. 

\emph{Orbital dependence of spin-spin correlations---}To investigate how different low-energy models affect the magnetic properties of LaNiO$_2$, we compute the local (impurity) spin–spin correlation function within the DFT+DMFT framework,
$\chi(\tau)$=$\langle S_z(\tau)\,S_z(0)\rangle$, where $S_z(\tau)$ denotes the $z$ component of the spin operator and $\tau$ is the imaginary time. While $\chi$($\tau$=0) reflects the instantaneous local magnetic moment, the value of $\chi$($\tau$) at $\tau$$\approx$$\beta$/2 characterizes the magnetic components that survive to long (imaginary) times.

\begin{figure}[t]
\centering
\includegraphics[width=\columnwidth]{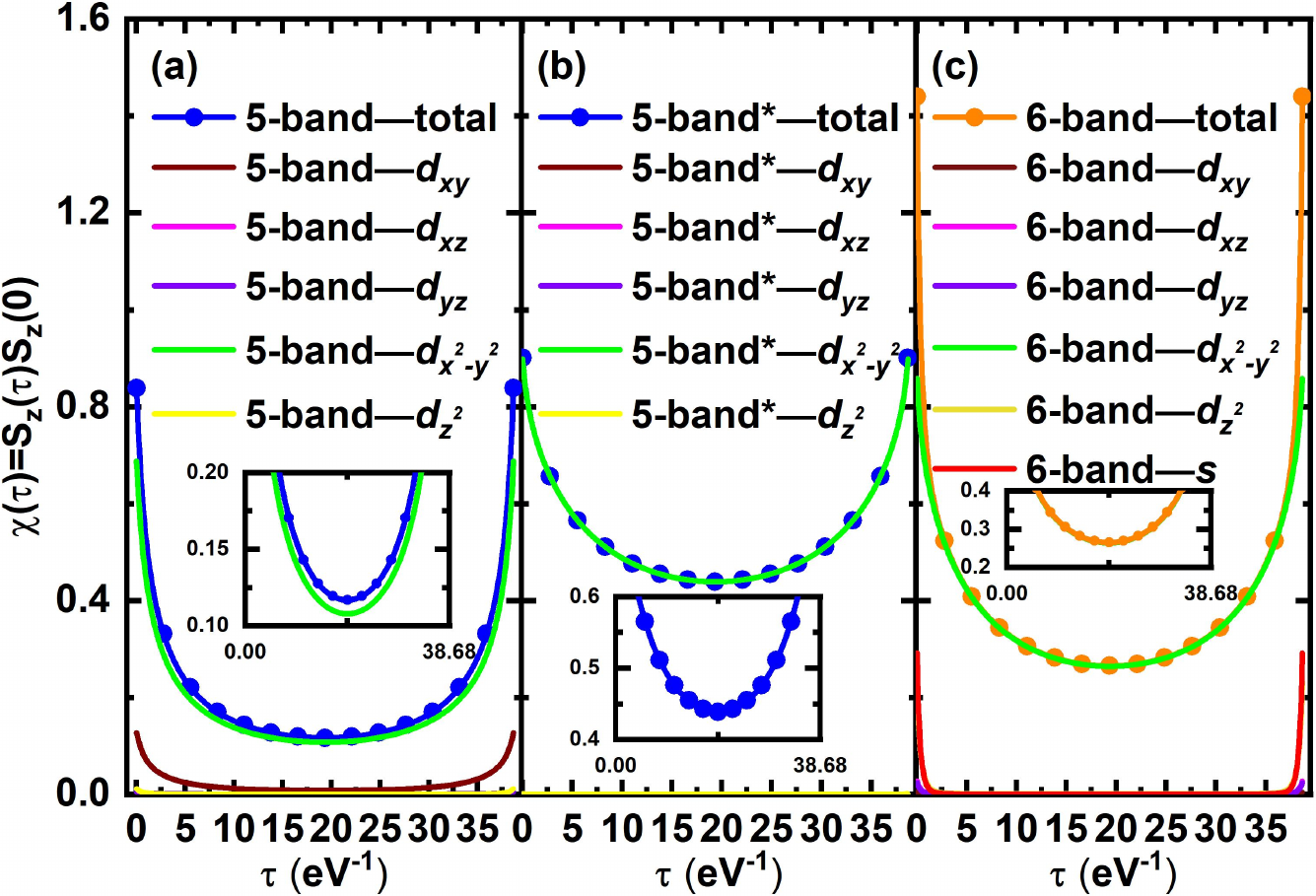}
\caption{Orbital-resolved local spin–spin correlation functions 
$\chi(\tau)=\langle S_z(\tau)\,S_z(0)\rangle$ of LaNiO$_2$ obtained from DMFT calculations at inverse temperature
$\beta=38.68$\,eV$^{-1}$ (corresponding to room temperature, 300\,K), for 
(a) 5-band, (b) filling-adjusted 5-band$^{\ast}$, and 
(c) 6-band model. The insets display a zoom-in of the main panel around 
$\tau\approx\beta/2$. In (c), in addition to the Ni-$3d$ orbital contributions, the interstitial-$s$ orbital component is also included.}
  \label{fig:figure4}
\end{figure}

Figure~\ref{fig:figure4}(a) shows the $\chi(\tau)$ for the 5-band model. At $\tau$=$\beta$/2, the total spin–spin correlation significantly exceeds the $d_{x^2-y^2}$ contribution, indicating that multiple Ni-3$d$ orbitals participate in the magnetic response, revealing a characteristic multi-orbital Hund-metal behavior. By contrast, in the 6-band model [Fig.~\ref{fig:figure4}(c)], the total $\chi$($\tau$) at $\tau$=$\beta$/2 coincides with the $d_{x^2-y^2}$ contribution, while all other $d$ orbitals are strongly suppressed, demonstrating a single-orbital-dominated magnetic response consistent with an orbital-selective Mottness. Although the interstitial-$s$ orbital exhibits a finite contribution at $\tau$=0, it decays rapidly with increasing $\tau$, indicating that it does not participate in long-time magnetic correlations.

To disentangle whether the distinct correlation regimes originate from the dimensionality of the low-energy Hamiltonian or from the effective filling of Ni-3$d$ orbitals, we further consider the filling-adjusted 5-band$^\ast$ model in Fig.~\ref{fig:figure4}(b).
The 5-band$^{\ast}$ model shows again a nearly pure $d_{x^2-y^2}$-dominated spin correlation pattern, closely resembling the 6-band model. These results demonstrate that the Hundness vs.~Mottness character is primarily governed by the effective filling of the Ni-3$d$ orbitals rather than by the dimensionality of the low-energy Hamiltonian or the explicit inclusion of the interstitial-$s$ orbital.

\emph{Conclusion---}In strongly correlated systems, increasing the number of sites or orbitals in models often improves credibility but dramatically raises computational cost. Determining the appropriate minimal model is therefore a central challenge, particularly for complex materials with multi-sites and multi-orbitals. Despite extensive DFT and DMFT studies in recent years, the minimal effective band model for nickelates has remained debated. Our calculations reveal two complementary modeling limits. First, due to the substantially weaker $d$-$p$ hybridization in nickelates compared with cuprates, the Ni-3$d_{x^2-y^2}$ orbital behaves increasingly like an isolated Hubbard band. When its filling is accurately determined using many-body methods such as DFT+DMFT, a doped 1-band $d_{x^2-y^2}$ model successfully reproduces experimentally observed superconducting domes. This establishes the 1-band Hubbard model as an adequate and efficient framework for describing superconductivity in infinite-layer nickelates \cite{kitatani2020nickelate,PhysRevLett.130.166002,PhysRevMaterials.6.L091801,PhysRevB.109.235126}. The 5-band model calculations significantly overestimate Hund’s metal behavior. Only if here the doping is also adjusted (5-band$^\ast$ model) is the correct  Mottness recovered. Only then is it qualitatively similar to the (also filling adjusted) 1-band and the 6-band models.
Second, for multi-band phenomena such as optical transitions, the simplified 1-band model becomes insufficient. Multi-band effects  are essential for high-energy optical properties
and the interstitial-$s$ band contributes substantially to the Drude weight. 

Our results demonstrate that \emph{the accuracy and efficiency of a minimal model cannot be judged solely by the number of orbitals}. Instead, the choice of model must be dictated by the physical properties under investigation. An inappropriate model can lead to qualitatively incorrect conclusions about single-band versus multi-band correlation physics.

\emph{Acknowledgments---}L.~S.~acknowledges support from the National Natural Science Foundation of China (Grant Nos.~12422407 and 12674309) and  the Key Research and Development Program of Shaanxi (2024QY2-GJHX-42).
Z.~X.~S.~and R.~Y.~acknowledge support from the National Natural Science Foundation of China (Grant Nos.~12374135 and 12474152) and the National Key R\&D Program of China (Grant No.~2024YFA1408400).
M.~J.~acknowledges the support of the National Natural Science Foundation of China (Grant No.~12174278).
Calculations have been done on the National Supercomputing Center in Northwest University (Xi'an) and the Vienna Scientific Cluster (VSC).

\emph{Data availability---}The data that support the findings of this article are openly available \footnote{Novel Materials Discovery (NOMAD) repository, \href{https://nomad-lab.eu/prod/v1/gui/user/datasets/dataset/id/AVsBc-9DSXWKtHj8TrR3XQ}{10.17172/NOMAD/2026.01.27-1}.}.

\clearpage
\onecolumngrid
\setcounter{figure}{0}
\setcounter{table}{0}
\setcounter{equation}{0}
\renewcommand{\thefigure}{S.\arabic{figure}}
\renewcommand{\thetable}{S.\Roman{table}}
\renewcommand{\theequation}{S.\arabic{equation}}
\renewcommand{\theHfigure}{supp.\arabic{figure}}
\renewcommand{\theHtable}{supp.\arabic{table}}
\renewcommand{\theHequation}{supp.\arabic{equation}}
\section*{Supplemental Material}
This Supplemental Material presents additional computational details and results from density-functional theory (DFT) and dynamical mean-field theory (DMFT) calculations that support the conclusions of the main text. Specifically, 
Section I compares different 1-band and 2-band Wannier projections of LaNiO$_2$. 
Section II provides DFT-level orbital-resolved band characters (fat-band analysis) of LaNiO$_2$. 
Section~III presents a real-space charge density visualization of the interstitial-$s$ band at $\Gamma$- and $A$-points, highlighting the momentum-dependent and three-dimensional spatial distribution of the  interstitial-$s$ pocket band.
Section~IV introduces a filling-adjusted 5-band$^\ast$ model
in which the chemical potential is tuned to restore a physically consistent Ni-$3d$ filling comparable to that of the 1-band and 6-band models. This construction provides a controlled reference for examining the role of effective electron filling in the 5-band description.
Section V provides additional DMFT spectral functions and orbital-resolved occupations. 
Section VI lists the tight-binding Hamiltonian hopping parameters obtained from Wannier projections for the 2-band and 6-band models, with particular emphasis on the interstitial-$s$ band and its orbital hopping to the $d_{x^2-y^2}$ band.
Section VII introduces the definition of the integrated optical conductivity and presents a quantitative analysis of optical spectral weight based on frequency-integrated conductivity.
Section VIII presents the extraction of quasiparticle weights and the comparison with ARPES measurements.

\section*{I.~WANNIER FUNCTION PROJECTIONS}
\begin{figure*}[h]
\centering
\includegraphics[width=\columnwidth]{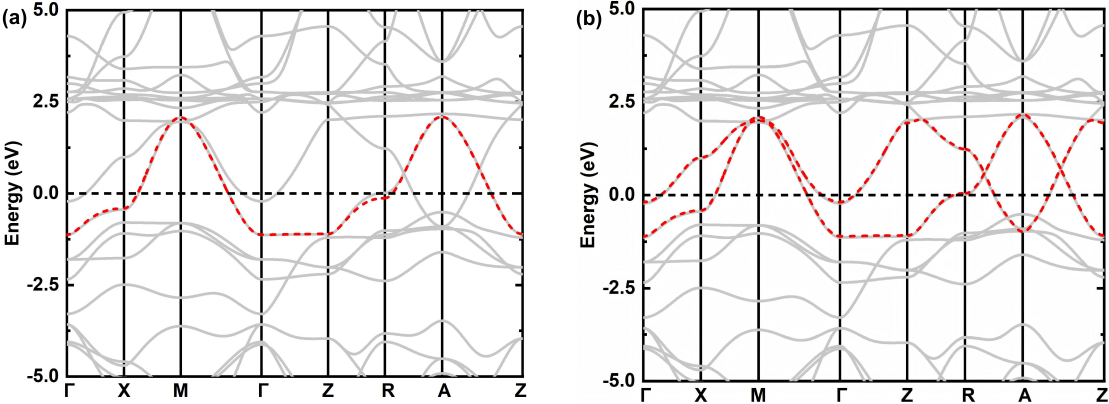}
\caption{Comparison of Wannier-projected band structures  (dashed red line) for (a) the 1-band (Ni-3$d_{x^2-y^2}$) model and (b) the 2-band (Ni-3$d_{x^2-y^2}$+interstitial-$s$) model overlaid with the DFT band structure (gray curves) of LaNiO$_2$.}
\label{fig:SM-fig1}
\end{figure*}

Figure~\ref{fig:SM-fig1} compares the Wannier-projected band structures of the 1-band and 2-band models with the first-principles DFT band structure of LaNiO$_2$. Both models (Wannier projections) reproduce the DFT bands well, demonstrating good consistency between the Wannier projections and the first-principles results.

\clearpage

\section*{II.~Orbital-resolved fat-band representation of the DFT band structure}
\begin{figure*}[h]
\centering  \includegraphics[width=0.8\columnwidth]{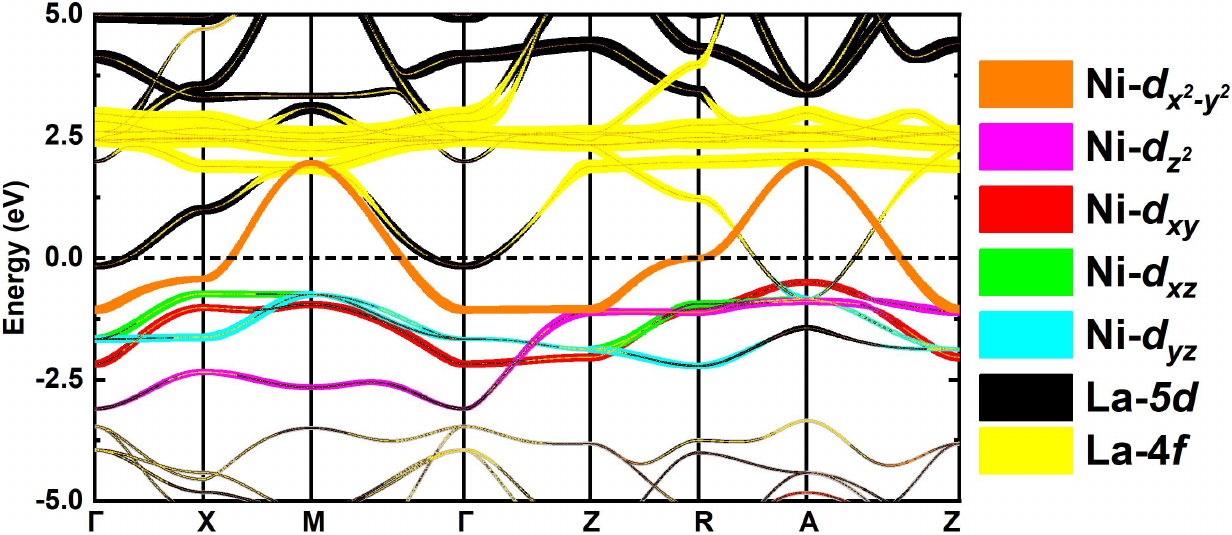}
\caption{Orbital-resolved DFT band structure (fat-band) of LaNiO$_2$. The first-principles DFT orbital-resolved electronic band structure is shown, where the thickness of the color lines reflects the weight of each orbital at a given energy and momentum $k$. Different colors indicate contributions from Ni-3$d$ orbitals, as well as La-5$d$ and La-4$f$ states.}
  \label{fig:SM-fig2}
\end{figure*}

Figure~\ref{fig:SM-fig2} presents the orbital-resolved (fat-band) band structure of LaNiO$_2$ along the same high-symmetry $k$-path as in Fig.~1 in the main text. The low-energy states near the Fermi level ($E_F$) are dominated by the Ni-3$d_{x^2-y^2}$ orbital, consistent with the quasi two-dimensional electronic structure induced by the NiO$_2$ planes in infinite-layer nickelates. This confirms that the Ni-3$d_{x^2-y^2}$ orbital governs the low-energy physics and the most simplified model of the system.

At the $\Gamma$-point, a pocket-like structure originates mainly from hybridization between Ni-3$d_{z^2}$ and La-5$d$ (predominantly 5$d_{z^2}$) states, highlighting its three-dimensional character. In addition, along $R$-$A$-$Z$, the $A$-pocket contains mixed contributions from La-4$f$, Ni-3$d_{yz}$, and Ni-3$d_{xz}$ orbitals. Both $\Gamma$- and $A$-pockets
can be included through an additional interstitial-$s$ orbital.

As discussed in the main text, at $A$-point the Ni-3$d_{yz}$ states are primarily located around -1\,eV rather than -2\,eV. This accounts for the noticeable mismatch between the DFT bands and the 5-band Wannier projection near A-point, underscoring the intrinsic limitations of the 5-band model in accurately capturing the DFT electronic band structure.

\clearpage

\section*{III.~Charge density of the interstitial-$s$ orbital}

\begin{figure*}[h]
\centering  \includegraphics[width=0.8\columnwidth]{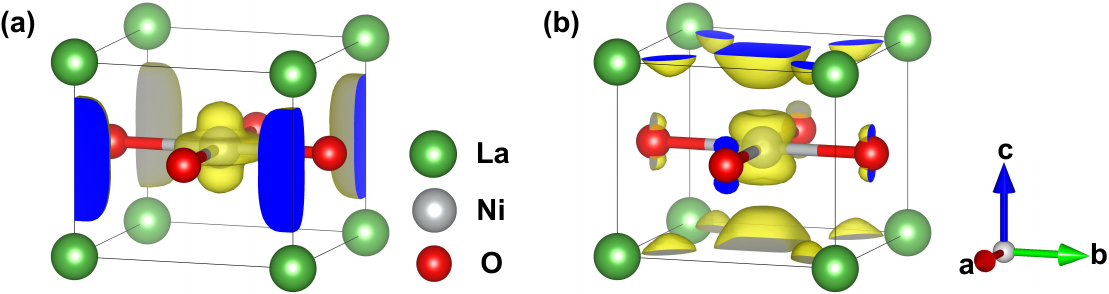}
\caption{Comparison of the real-space distributions of interstitial-$s$ states at the $\Gamma$-point and near the $A$-point.}
  \label{fig:SM-fig3}
\end{figure*}

To further investigate the origin of the interstitial-$s$ orbital and its contribution to the low-energy electronic structure at different momenta $k$, we visualize the real-space charge density of the interstitial-$s$ band at the $\Gamma$- and $A$-points at the DFT level. As shown in Fig.~\ref{fig:SM-fig3}, the charge density exhibits a pronounced momentum dependence. At the $\Gamma$-point, the interstitial-$s$ band displays a quasi-one-dimensional character along the $z$-direction, arising primarily from hybridization between La-5$d_{z^2}$ and Ni-3$d_{z^2}$ orbitals.
In contrast, near the $A$-point the charge density exhibits a pronounced interstitial character, with substantial weight distributed in the regions between the NiO$_2$ layers and a clear hybridization with La-$5d$ states. The resulting spatial distribution is strongly three-dimensional and extended along the $c$ axis, reflecting the highly itinerant nature of the interstitial-$s$–derived electronic states.

\clearpage

\section*{IV.~Filling-adjusted 5-band reference (5-band$^*$)}

In the main text, we compared the electronic structures and correlation effects of LaNiO$_2$ using different low-energy effective models and found that the 5-band model containing only Ni-3$d$ orbitals and the 6-band model including both Ni-3$d$ and interstitial-$s$ orbitals exhibit qualitatively distinct behaviors. The former is more consistent with a Hund-type multi-orbital correlated metal, whereas the latter is closer to an orbital-selective Mottness picture dominated by the $d_{x^2-y^2}$ orbital. To determine whether this distinction arises from the dimensionality of the low-energy model itself or from differences in effective filling and charge redistribution, we introduce a filling-adjusted 5-band reference model, denoted as 5-band$^*$ model.

In multi-band models that explicitly include La-5$d$ and interstitial-$s$ orbitals—such as the La-5$d$+Ni-3$d$ (10-band model) or the La-5$d$+Ni-3$d$+interstitial-$s$ (11-band model)—these additional states hybridize with the Ni-3$d$ bands near and below the Fermi level E$_f$, contributing finite spectral weight below E$_f$.
Due to the $A$- (and $\Gamma$-)pocket there is a self-doping in infinite-layer nickelates. The Ni-$3d$ occupation can therefore differ from the nominal $d^9$ electron count. Also the aforementioned hybridizations can change the occupation of the 10- or 11-band models; however the  hybridization to states above and below the Fermi energy largely cancels, see Ref.~\cite{hausoel2025superconducting}.

To overcome this band filling effect, we construct the 5-band$^*$ model using exactly the same low-energy Hamiltonian and interaction parameters as in the original 5-band model, but with the chemical potential $\mu$ adjusted such that the $d_{x^2-y^2}$ orbital occupation is restored to $n_{x^2-y^2}$=0.95, consistent with both the 6-band model and the 1-band model calculations.

\begin{figure*}[h]
\centering
\includegraphics[width=0.8\columnwidth]{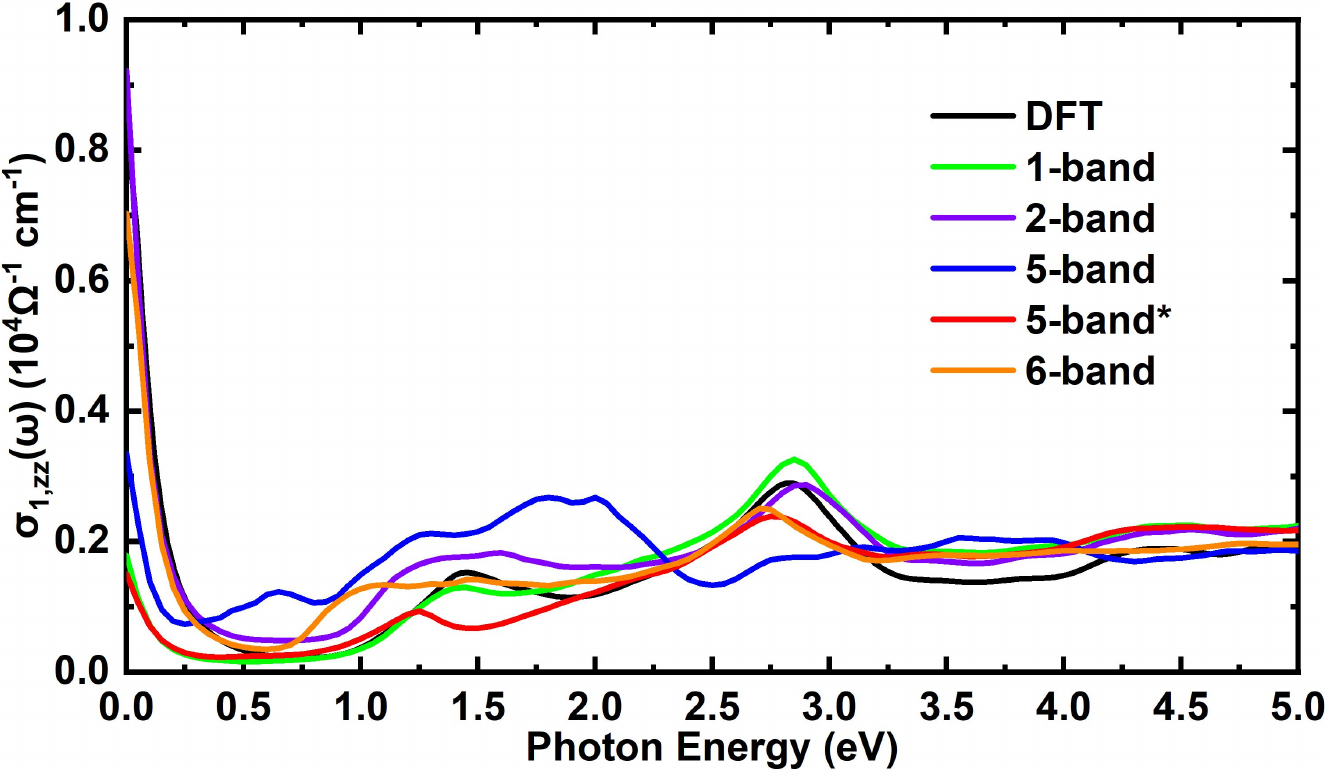}
\caption{Optical conductivity $\sigma(\omega)$ of LaNiO$_2$ computed with different effective models, including the 1-band, 2-band, 5-band, filling-adjusted 5-band (5-band$^*$), and 6-band models. (Uncorrelated) DFT $\sigma(\omega)$ is also plotted for comparison.}
  \label{fig:SM-fig4}
\end{figure*}

Figure~\ref{fig:SM-fig4} compares the optical conductivity $\sigma$($\omega$) of the 5-band, filling-adjusted 5-band (5-band$^*$), and 6-band models. While the unadjusted 5-band model exhibits a pronounced absorption peak around $\sim$2\,eV, this feature is strongly suppressed and nearly absent in the 5-band$^*$ calculation. Correspondingly, the overall spectral shape of the 5-band$^*$ model becomes much closer to that of the 6-band model in the same energy range, and also approaches the single-band result in the low-frequency Drude regime. This demonstrates that the prominent mid-energy absorption peak in the 5-band model is not an intrinsic property of the 5-band Hamiltonian itself, but rather originates from the effective filling and the enhanced Hund-type correlations present in the unadjusted 5-band calculation. When the $d_{x^2-y^2}$ orbital is restored to near half-filling, the multi-orbital inter-band excitations responsible for this peak are strongly suppressed, and the optical response becomes characteristic of a Mottness or orbital-selective correlated metal dominated by a single low-energy orbital.

\begin{figure*}[h]
\centering
\includegraphics[width=0.8\columnwidth]{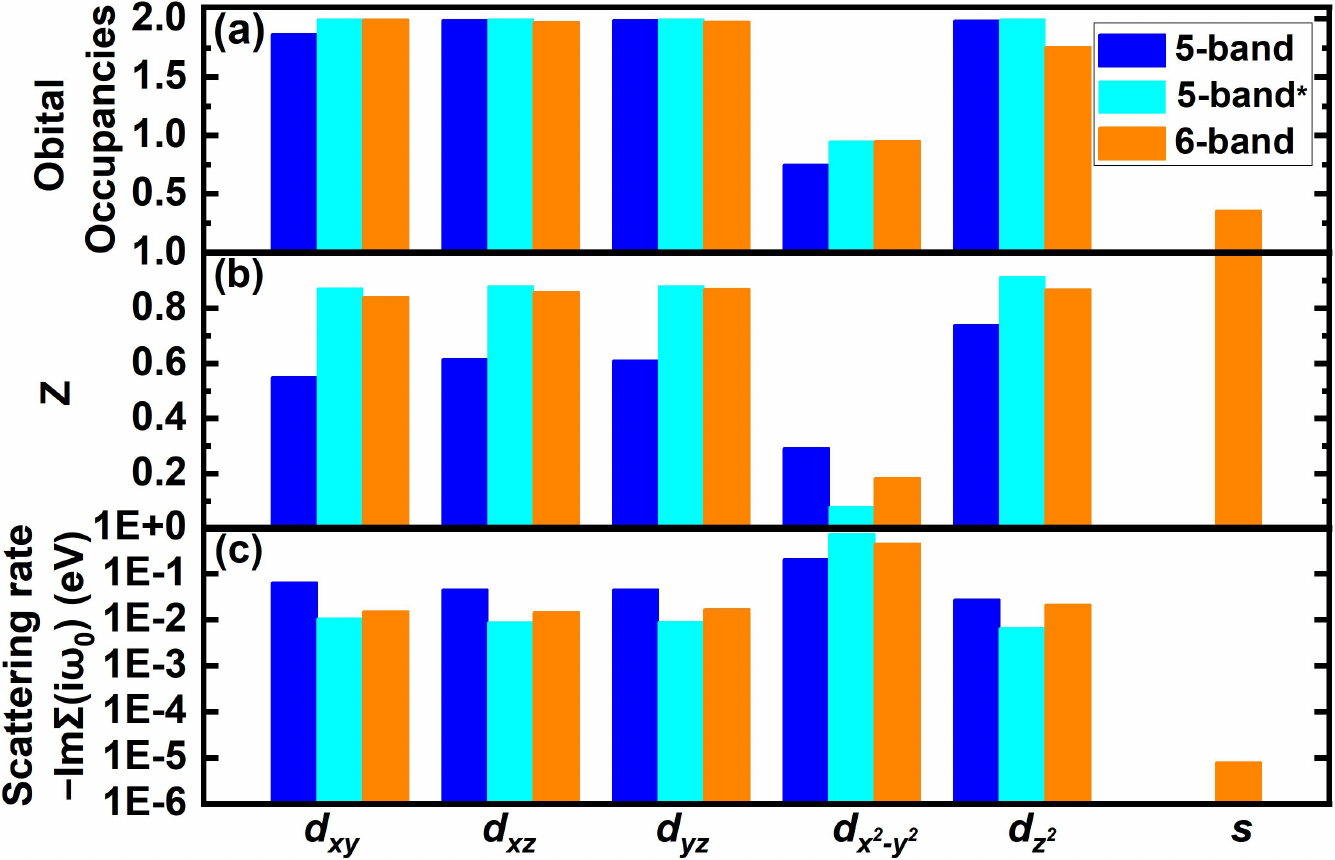}
\caption{(a) Orbital-resolved Ni-$3d$ occupations, (b) quasi-particle renormalization factors $Z$, and (c) scattering rates $-\mathrm{Im}\,\Sigma(i\omega_0)$ for different low-energy models, including a comparison between the 5-band and the filling-adjusted 5-band (5-band$^*$) calculations.}
  \label{fig:SM-fig5}
\end{figure*}

Figure~\ref{fig:SM-fig5} further demonstrates the effect of filling adjustment at DMFT level. The orbital occupations [Figure~\ref{fig:SM-fig5}(a)] confirm that in the 5-band$^*$ model the $d_{x^2-y^2}$ orbital is tuned to $n$($d_{x^2-y^2}$)=0.95, consistent with the 6-band model and enabling a meaningful comparison. With this adjustment, the quasi-particle renormalization factors $Z$ and the low-frequency scattering rates $-\mathrm{Im}\,\Sigma(i\omega_0)$ reveal a clear crossover from Hundness correlations to orbital-selective Mottness. In the 5-band$^*$ model, the $d_{x^2-y^2}$ orbital exhibits substantially stronger renormalization (smaller $Z$) and a markedly enhanced scattering rate, while the remaining Ni-3$d$ orbitals are only weakly renormalized. This results in a strongly orbital-selective correlation pattern that closely mirrors the behavior of the 6-band model. By contrast, in the unadjusted 5-band model all $d$ orbitals contribute more uniformly to the low-energy renormalization and scattering, characteristic of a Hund-metal.

Taken together, Figs.~\ref{fig:SM-fig4}–\ref{fig:SM-fig5} show that, once the filling of the 5-band model is adjusted to reproduce $n(d_{x^2-y^2})$=0.95, its optical response and single-particle correlation properties—namely the quasi-particle renormalization factors and scattering rates—become quantitatively closer to those of the 6-band model. In particular, the pronounced suppression of the mid-energy optical peak and the enhanced orbital selectivity in the renormalization and scattering behavior indicate that the effective electron filling plays the decisive role in shaping the correlation regime. The filling-adjusted 5-band$^*$ model therefore serves as a controlled reference that disentangles the impact of electron occupation from the dimensionality of the low-energy Hamiltonian. These supplementary results provide additional support for the filling-driven evolution of correlation behavior discussed in the main text.

\clearpage

\section*{V.~DMFT spectral functions with inclusion of $U$(interstitial-$s$) and various band fillings}

\begin{figure*}[h]
\centering
\includegraphics[width=0.7\columnwidth]{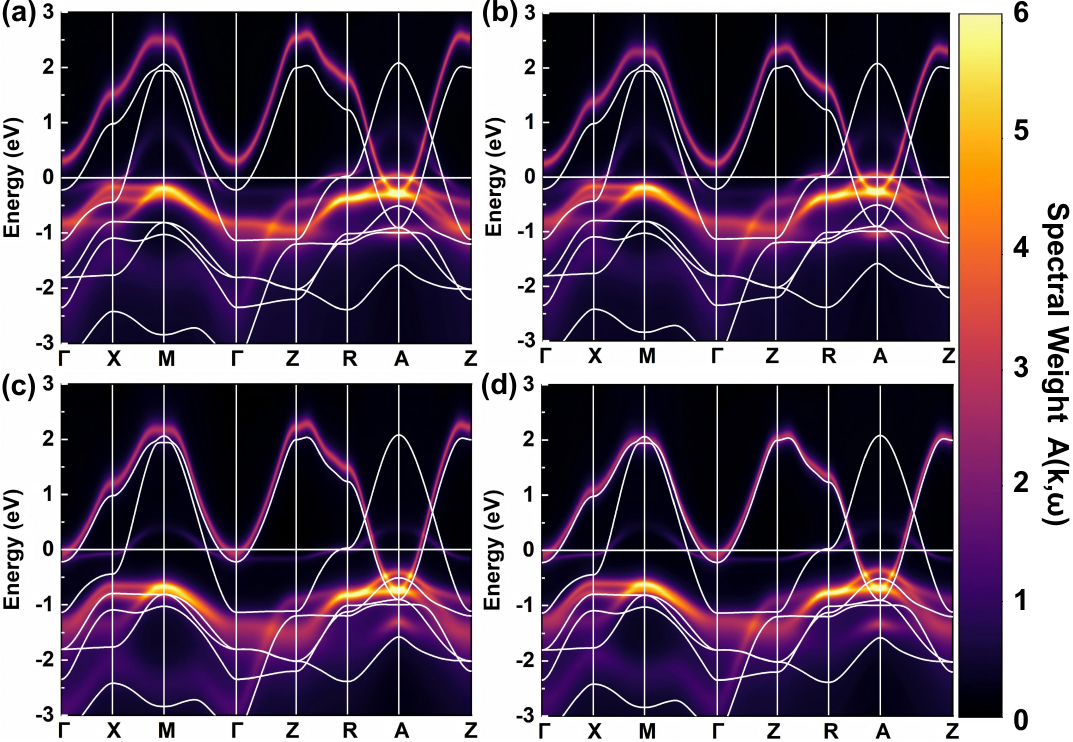}
\caption{Momentum-resolved DMFT spectral functions $A$($\omega$,$\mathbf{k}$) for the 6-band model with (a,b) and without (c,d) Sr doping, without (a,c) and with (b,d) 
interaction for the interstitial-$s$ orbital. Specifically: (a) $U$(interstitial-$s$)=0.0\,eV and $n$=8.8 (for simulating 20\% Sr doping); (b) $U$(interstitial-$s$)=2.5\,eV and $n$=8.8; (c) $U$(interstitial-$s$)=0.0\,eV and $n$=9.0, and (d) $U$(interstitial-$s$)=2.5\,eV and $n$=9.0.}
  \label{fig:SM-fig6}
\end{figure*}

In previous theoretical and experimental studies of infinite-layer nickelate superconductors, a central issue is whether an electronic pocket exists at the $\Gamma$-point in the undoped and (Sr-induced) hole-doped samples. In the main text, we adopted a nominal Ni$^+$ valence with a total electron count of $n$=9.0 within the 6-band model and treated the interstitial-$s$ orbital as uncorrelated. The resulting spectral function [Fig.~\ref{fig:SM-fig6}(c) and main text Fig. 1(d)] shows a clear $\Gamma$-pocket. An important question is whether this pocket survives upon 20\% hole doping ($n$=8.8), especially given that the interstitial-$s$ orbital contains La-5$d$ and Ni-3$d$ character and may therefore be subject to correlation effects. Including correlations on the interstitial-$s$ orbital could shift this band upward and potentially remove the $\Gamma$-pocket. To address this issue, we repeated the 6-band DMFT calculations under different conditions [Fig.~\ref{fig:SM-fig6}(a-d)]. For $n$=8.8 [Fig.~\ref{fig:SM-fig6}(a,b)], no 
$\Gamma$-point pocket is found, irrespective of whether the interstitial-$s$ orbital is treated as correlated or uncorrelated. In contrast, for $n$=9.0 [Fig.~\ref{fig:SM-fig6}(c,d)], the $\Gamma$-point pocket remains robust in both cases.

Table~\ref{tabs:occupation} summarizes the orbital occupations obtained from the DMFT calculations for the 6-band model shown in Fig.~\ref{fig:SM-fig6}. The results show that hole doping predominantly affects the Ni-3$d_{x^2-y^2}$ orbital, whose occupation changes by approximately 0.12 electrons, while the occupations of all other orbitals vary only slightly, by about $\sim$0.02 electrons. This demonstrates that the essential low-energy physics of the system is governed by a single-band Mott scenario dominated by the $d_{x^2-y^2}$ orbital.

\begin{table*}[h]
\centering
\caption{DMFT computed orbital-resolved electron occupations in the 6-band model for fillings $n$=8.8 and $n$=9.0 and an interstitial-$s$ interaction $U$ of 0.0\,eV and 2.5\,eV.}
\label{tabs:occupation}
\setlength{\tabcolsep}{10pt}
\begin{tabular}{c c c c c c c}
\hline\hline
Band filling $n$ and Coulomb $U$ & Ni-$d_{xy}$ & Ni-$d_{xz}$ & Ni-$d_{yz}$ & Ni-$d_{x^2-y^2}$ & Ni-$d_{z^2}$ & $s$\\
\hline
$n$=8.8; $U$=0.00\,eV
 & 1.97 & 1.96 & 1.96 & 0.83 & 1.73 & 0.35 \\
\hline
$n$=8.8; $U$=2.50\,eV
 & 1.97 & 1.96 & 1.96 & 0.83 & 1.71 & 0.37 \\
 \hline
$n$=9.0; $U$=0.00\,eV
 & 1.99 & 1.97 & 1.97 & 0.95 & 1.76 & 0.36 \\
 \hline
$n$=9.0; $U$=2.50\,eV
 & 1.99 & 1.97 & 1.97 & 0.94 & 1.74 & 0.38 \\
 
\hline\hline
\end{tabular}
\vspace{2mm}
\end{table*}

\clearpage

\section*{VI.~Effective Hamiltonian and hopping parameters from Wannier projections}

Table~\ref{tab:onsite} lists the on-site terms [i.e., hoppings at R=(0,0,0)] for different orbitals obtained from Wannier projections of the various models relative to the Fermi energy $E_F$=0. The on-site energy of the Ni-3$d_{x^2-y^2}$ orbital is nearly identical in the 1-, 2-, 5-, and 6-band models, indicating that this orbital is effectively decoupled from the others.  In particular, the on-site hybridization between Ni-3$d_{x^2-y^2}$ and other orbitals is essentially zero due to orbital orthogonality; a similar situation holds for the Ni-3$d$ $t_{2g}$ orbitals. However, two notable exceptions appear: First, the on-site energy of the Ni-3$d_{z^2}$ orbital differs by about 450\,meV between the 5- and 6-band models, arising from direct hybridization with the interstitial-$s$ orbital along the (001) $z$-direction. This leads to a shift of the bonding Ni-3$d_{z^2}$ orbital to lower energies in the 5-band model where this hybridization is no longer explicitly included. Second, in the 2- and 6-band models, the interstitial-$s$ orbital on-site energy varies significantly, reflecting its strong overlap with other orbitals and highlighting the nonlocal character of the interstitial-$s$ orbital.

\begin{table*}[h]
\centering
\caption{On-site hopping terms (in units of meV) in the corresponding tight-binding Hamiltonians for different low-energy models.}
\label{tab:onsite}
\setlength{\tabcolsep}{10pt}
\begin{tabular}{c|c|c|c|c|c|c|c}
\hline
\hline
Model
& Orbital content
& {Ni-3$d_{xy}$}
& {Ni-3$d_{xz}$}
& {Ni-3$d_{yz}$}
& {Ni-3$d_{x^2-y^2}$}
& {Ni-3$d_{z^2}$}
& {Interstitial-$s$} \\
\hline
1-band &Ni-3$d_{x^2-y^2}$ &  {--} & {--} & {--} & 262.5 & {--} & {--} \\
\hline
2-band
& Ni-3$d_{x^2-y^2}$+interstitial-$s$ & {--} & {--} & {--} & 289.5 & {--} & 108.4\\
\hline
5-band
& Ni-$3d$ & -1281.7 & -1460.5 & -1460.5 & 257.0 & -2002.3 & {--}\\
\hline
6-band
& Ni-$3d$+interstitial-$s$ & -1284.9 & -1419.0 & -1419.0 & 262.5 & -1554.0 & 561.5 \\
\hline
\hline
\end{tabular}
\end{table*}

Table~\ref{tab:s_d_hopping} lists the hopping terms between the interstitial-$s$ orbital and the Ni-3$d_{x^2-y^2}$ orbital, leading to the following observations: For the R=(0,0,0) (on-site) and (0,0,1) (out-of-plane direction) directions, the interstitial-$s$ to $d_{x^2-y^2}$ hoppings are (nearly) zero, reflecting the orthogonality between these two orbitals; along the (1,0,0) direction, the interstitial-$s$ to $d_{x^2-y^2}$ hoppings are up to 50\,meV and 252\,meV in the 6- and 11-band models, respectively. Along the (1,1,0) direction, the interstitial-$s$ to $d_{x^2-y^2}$ hopping is again nearly zero, protected by orbital symmetry. For the (2,0,0) direction, both interstitial-$s$ to interstitial-$s$ and interstitial-$s$ to $d_{x^2-y^2}$ hoppings are on the order of meV to 10\,meV, which can be attributed to the nonlocal character of the interstitial-$s$ orbital.

\begin{table*}[h]
\centering
\caption{Wannier hopping parameters (in units of meV) between the interstitial-$s$ orbital and to Ni-$d_{x^2-y^2}$ orbital for the 6-band and 11-band models. The 6-band model consists of the five Ni-$3d$ orbitals together with the interstitial-$s$ orbital, while the 11-band model further includes the five La-$5d$ orbitals in addition to the Ni-$3d$ and interstitial-$s$ orbitals. The hopping directions $t_{000}$, $t_{100}$, $t_{001}$, $t_{110}$, and $t_{200}$ correspond to on-site (000), the 1st nearest neighbor along the in-plane (100), the out-of-plane direction along (001), 2nd  nearest neighbor along (110), and the 3rd nearest neighbor hopping along (200), respectively.}
\label{tab:s_d_hopping}
\setlength{\tabcolsep}{10pt}
\begin{tabular}{c|c|c|c|c|c|c}
\hline\hline
Model & Orbitals & $t_{000}$ & $t_{100}$ & $t_{001}$ & $t_{110}$ & $t_{200}$  \\
\hline
\multirow{2}{*}{6-band}
 & $s-s$ & 561.5 & -11.7 & -76.5 & -137.4 & 9.1 \\
 & $d_{x^2-y^2}-s$ & 4.1 & 50.0 & 4.3 & -3.1 & 20.2 \\
\hline
 \multirow{2}{*}{11-band}
 & $s-s$ & 327.0 & -99.7 & 102.3 & 170.3 & 4.4 \\
 & $d_{x^2-y^2}-s$ & 0.0 & 252.2 & 0.0 & 0.0 & 9.6 \\
\hline\hline
\end{tabular}
\vspace{2mm}
\end{table*}

\clearpage

\section*{VII.~Definition and comparison of integrated optical conductivity}
\begin{figure*}[h]
\centering
\includegraphics[width=0.8\columnwidth]{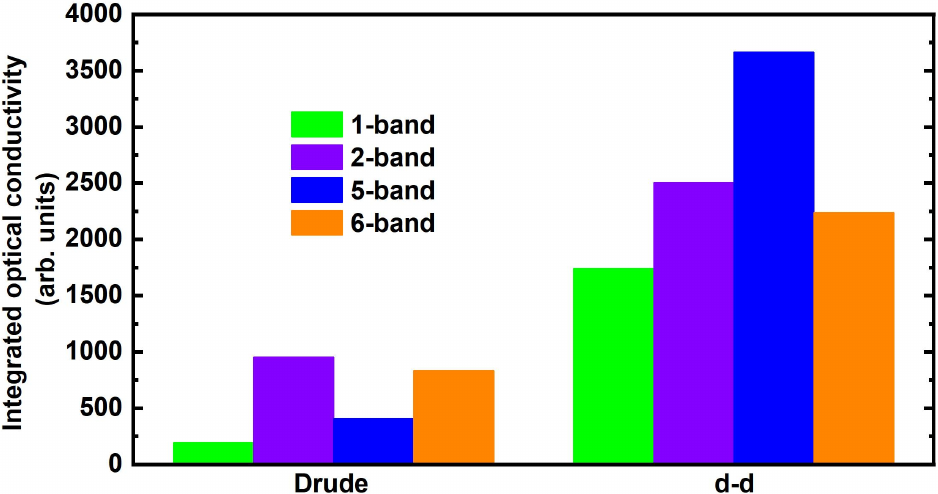}
\caption{Integrated optical spectral weight. Comparison of the integrated optical conductivity for different low-energy models (1-band, 2-band, 5-band, and 6-band). The left bar chart shows the low-frequency Drude weight, while the right bar chart displays the integrated spectral weight associated with mid-infrared $d$-$d$ transitions. The integration windows are defined consistently with those used in the main text.}
  \label{fig:SM-fig7}
\end{figure*}

To quantitatively assess how optical responses are distributed across different energy scales in the various low-energy effective models, we integrate the real part of the optical conductivity (Fig.~\ref{fig:SM-fig7}), $\sigma_1$($\omega$), over selected frequency windows. This analysis separates low-energy carrier contributions from intermediate- and high-energy inter-orbital transitions, enabling a clear comparison of electronic dynamics and multi-orbital effects among the models.

Specifically, the low-frequency Drude weight is defined as:
\[
W_{\mathrm{Drude}}
=
\int_{0}^{\omega_c} \sigma_1(\omega)\, d\omega,
\]

where the cutoff frequency $\omega_c = 0.296\,\mathrm{eV}$ is chosen to exclude pronounced interband transition contributions. Thus, this integral primarily reflects the contribution of low-energy quasiparticles to the optical response. This Drude weight is commonly associated with effective mass renormalization, carrier density, and low-energy coherence.

In addition, to characterize the spectral-weight distribution in the mid-infrared energy range dominated by optical transitions between different $d$ orbitals, we define
\[
W_{d\text{-}d}
=
\int_{\omega_1}^{\omega_2} \sigma_1(\omega)\, d\omega,
\]

where the integration window $[\omega_1,\omega_2]=[0.296,2.3]\,\mathrm{eV}$ spans the energy range dominated by $d$-$d$ transitions. The resulting integrated quantity thus characterizes the strength of multi-orbital effects and inter-orbital transitions in the optical response. The specific integration windows are the same as those of the main text for the analysis of the optical conductivity spectra.

\clearpage

\section*{VIII.~Quasiparticle renormalization and comparison with ARPES}
\begin{figure*}[h]
\centering
\includegraphics[width=0.8\columnwidth]{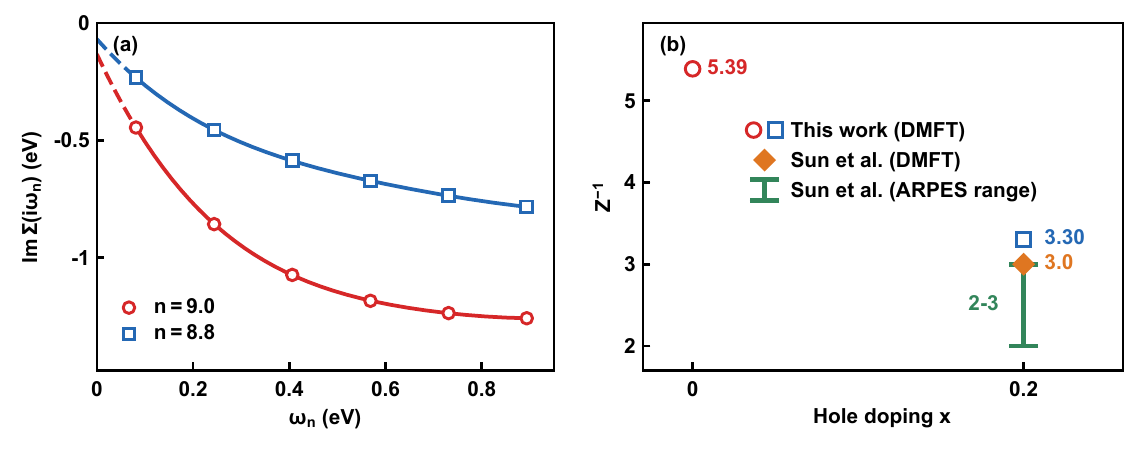}
\caption{Low-frequency self-energy fits and comparison with ARPES.
(a) Imaginary part of the Ni-3$d_{x^2-y^2}$ self-energy in the 6-band model at $n$=9.0 and $n$=8.8.
(b) Inverse quasiparticle weights obtained from the fitted zero-frequency slopes, compared with the DMFT estimate and ARPES renormalization range reported by Sun \textit{et al.} ~\cite{sun2025electronic}.}
  \label{SM-fig8}
\end{figure*}

To assess the quasiparticle renormalization and its dependence on the extraction procedure, we compare two different computational ways of the
quasiparticle weight (renormalization factor) $Z$ for the Ni-3$d_{x^2-y^2}$ orbital in the 6-band model. We consider total fillings of $n$=9.0 and $n$=8.8, corresponding to nominal hole dopings of $x$=0 and $x$=0.2 per unit cell, respectively. 

The real-frequency procedure used for the 6-band results in Fig.~3 of the main text estimates the quasiparticle weight from
the zero-frequency derivative of the real part of the analytically continued self-energy:
\begin{equation}
Z_{\mathrm{real}}^{-1}
=
1-
\left.
\frac{\partial\operatorname{Re}\Sigma(\omega)}
{\partial\omega}
\right|_{\omega=0}.
\label{eq:Z_real}
\end{equation}
This procedure yields $Z_{\mathrm{real}}^{-1}\approx$5.44 at $n$=9.0 and $Z_{\mathrm{real}}^{-1}\approx$4.05 at $n$=8.8.

As an additional check, we follow the Matsubara-frequency fitting procedure employed by Sun \textit{et al.}~\cite{sun2025electronic}. The imaginary part of the spin-averaged self-energy at the first six positive Matsubara frequencies is fitted using an unweighted fourth-order polynomial with a free constant term:
\begin{equation}
\operatorname{Im}\Sigma(i\omega_n)
\approx
a_0+a_1\omega_n+a_2\omega_n^2+a_3\omega_n^3+a_4\omega_n^4.
\label{eq:selfenergy_polynomial}
\end{equation}
The inverse quasiparticle weight is then estimated as
\begin{equation}
Z_{\mathrm{fit}}^{-1}
=
1-
\left.
\frac{\partial\operatorname{Im}\Sigma(i\omega)}
{\partial\omega}
\right|_{\omega\rightarrow0^+}.
\label{eq:Z_fit}
\end{equation}
At $\beta=38.68~\mathrm{eV}^{-1}$, these six frequencies span approximately 0.081--0.893\,eV. The self-energy data and corresponding fits are shown in Fig.~\ref{SM-fig8}(a). We obtain $Z_{\mathrm{fit}}^{-1}\approx5.39$ at $n$=9.0 and
$Z_{\mathrm{fit}}^{-1}\approx3.30$ at $n$=8.8.

The two procedures give closely similar results at $n$=9.0, whereas a slight difference is found at $n$=8.8.
Both nevertheless indicate that hole doping reduces the quasiparticle renormalization. The numerical estimates at finite temperature depend on the extraction procedure: the real-frequency derivative is sensitive to analytic continuation and the finite-difference step, whereas the Matsubara estimate depends on the polynomial order and fitting window.

Figure~\ref{SM-fig8}(b) compares the Matsubara-fit estimates with the results reported by Sun \textit{et al.}~\cite{sun2025electronic}. Their ARPES measurements on La$_{0.8}$Sr$_{0.2}$NiO$_2$ yielded a low-energy band renormalization of approximately 2--3, depending on the momentum cut, while their accompanying DMFT calculation gave a mass enhancement of approximately 3.0. At the corresponding nominal doping, both of our estimates (4.05 and 3.30) are closer to experiment than their respective undoped values, although both remain above the reported ARPES range. Here, comparison of $Z^{-1}$ with the experimental mass enhancement uses the dominant-orbital, local-self-energy approximation.

A direct quantitative comparison is further limited by differences in temperature, orbital space, and interaction parameters.
In particular, the ARPES measurements and the accompanying DMFT calculations were performed at lower temperatures than our calculations. Using the same number of Matsubara points and polynomial order therefore does not imply the same fitting
frequency window. These results support a doping-induced reduction of quasiparticle renormalization while also illustrating the method dependence of its finite-temperature numerical estimate.

\begin{figure*}[h]
\centering
\includegraphics[width=0.8\columnwidth]{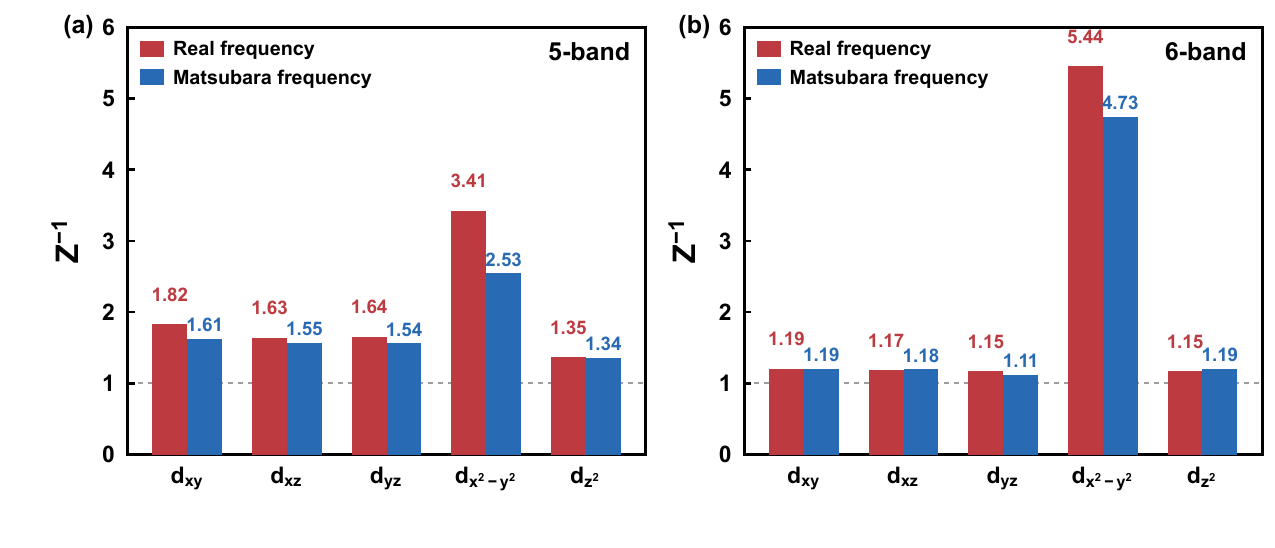}
\caption{Comparison of orbital-resolved inverse quasiparticle weights $Z^{-1}$ extracted from real-frequency and Matsubara-frequency self-energies for (a) the 5-band model and (b) the 6-band model. The gray dashed line indicates $Z^{-1}=1$.}
  \label{SM-fig9}
\end{figure*}

The above comparison shows that quantitative estimates of quasiparticle renormalization depend on the self-energy analysis procedure. To examine whether this dependence affects the comparison between the 5-band and 6-band models in the main text, Fig.~\ref{SM-fig9} presents the orbital-resolved inverse quasiparticle weights $Z^{-1}$ obtained from real-frequency and Matsubara-frequency self-energies for both models. The real-frequency estimates are obtained from the zero-frequency derivative of the real part of the self-energy, according to Eq.~(S.1). As an additional check, the Matsubara-frequency estimates shown here are extracted from the zero-frequency slopes of quadratic polynomial fits to the imaginary part of the self-energy over $0<\omega_n\leq0.5$~eV. Although the two procedures yield different numerical values of $Z^{-1}$, they show similar orbital dependence and trends across the two models, indicating that the main physical trends are consistent between the two extraction procedures.

\clearpage
\makeatletter
\let\auto@bib@innerbib\@empty
\makeatother

\end{document}